\documentclass[10pt]{article}
\title{Three Tests of General Relativity via Fermat's
Principle and the Phase of Bessel Functions}
\author{B.~H.~Lavenda\thanks{e-mail:
bernard.lavenda@unicam.it Work supported, in part, by
MIUR 60 \%.}\\ Universit\`a degli Studi\\
Camerino 62032 (MC) Italy}
\date{}

\usepackage{amsfonts}

\usepackage{eufrak}
\newcommand{\half}{\mbox{\small$\frac{1}{2}$}}

\newcommand{\fourth}{\mbox{\small$\frac{1}{4}$}}

\newcommand{\threehalves}{\mbox{\small$\frac{3}{2}$}}
\newcommand{\threefourths}{\mbox{\small$\frac{3}{4}$}}
\newcommand{\ninehalves}{\mbox{\small$\frac{9}{2}$}}

\begin{document}
\maketitle

\begin{abstract}
Fermat's principle applied to a flat metric in the plane
yields the phase of a Bessel function in the periodic domain
for a constant index of refraction. Gravitational forces
cause the index of refraction to vary and lead to a modified
phase of the Bessel function.  A distinction is made
between the forces that cause acceleration: the
gravitational force affects the optical properties of the
medium whereas the centrifugal force does not, the latter
being built into the phase of oscillations of the Bessel
function.  The time delay in radar echoes from planets is
determined from Fermat's principle where the velocity of
light is the phase velocity and the index of refraction
varies on account of the gravitational potential. The
deflection of light by a massive body is shown to be
produced by a quadrupole interaction, and the perihelion
shift requires both the gravitational potential, producing a
closed orbit, and the quadrupole, causing the perihelion to
rotate.

\end{abstract}

\vspace{5pt}
\section{Introduction}
General relativity is based on the notion that gravity,
rather than being a force between masses, is the curvature of
spacetime. The source of curvature is mass itself, like the
source of an electric field is charge. Just as free
particles follow   straight lines  in flat
spacetime, free particles  follow geodesics in the curved
spacetime of a gravitational field. The calculations of the
red shift, the time delay in radar echoes from planets,the
bending of light,   the perihelion shift, and the geodesic
effect all support general relativity.\par Sometimes new
insight can be gained in looking at old results from a new
perspective. The old quantum theory of multiply-periodic
central motion was superseded by wave mechanics. Old quantum
theory attempted to apply celestial mechanics to the world
of atoms with the addition of quantum restrictions to the
motion
\cite{Born}. In this article we will apply it to diffraction
phenomena in the short-wavelength limit that involve the
relativistic effects of time delay in radar sounding, the
deflection of light, and the advance of the
perihelion---three of the so-called five  tests of general
relativity.  Moreover, these relativistic effects are both
static  so that they can be investigated
  by studying only the spatial component of the metric.
The trajectory of a light ray in a static gravitational
field is determined in the same way as in an inhomogeneous
refractive medium \cite{Fock}. Fermat's principle of least
time relates the length and orientation of a light ray to the
time for light to propagate along the ray path.  The analogy
between the index of refraction and the square root of twice
the difference between the total and potential energies is
also well known. Yet, this can provide a way of
distinguishing between the centrifugal and gravitational
fields which cause acceleration, and  the form of the
trajectory of the motion.
\par   If we take a flat space time metric in the plane, and
consider a constant index of refraction, we will show that
Fermat's principle yields precisely the phase of the
oscillations of the Bessel function of the first kind in the
periodic domain in the asymptotic, short-wavelength, limit.
This is the same as the WKB result, and it allows us to
associate a wave phenomenon with a geodesic trajectory. The
only potential appearing is the repulsive centrifugal
potential, and the trajectory is necessarily open. In
contrast to general relativity, where gravity is not
considered as a force in the conventional sense, but,
rather, built into the curvature of spacetime, the
centrifugal force is built into   the phase of the Bessel
function in the periodic domain, where the trajectory
consists of straight line segments and arc lengths on a
caustic circle whose radius is determined by the magnitude
of the angular momentum. Whereas all forces causing
acceleration  are on the same
footing  in general relativity, the
gravitational force, and not centrifugal force, has to be
introduced through a space varying index of refraction in an
inhomogeneous refractive medium. The effect of gravity is to
make the medium optically more dense in the vicinity of a
massive body, while the centrifugal force has no effect on
the optical properties of the medium. Centrifugal, Coriolis
and gravitational forces are usually considered to be
fictitious insofar as they can be transformed away by a
change of frame. The centrifugal and Coriolis forces can be
transformed away by changing to a nonrotating frame, while
the fictitious force of gravity is transformed away by
changing from a nonfreely falling  to a freely falling
frame. In Newtonian theory one insists on a nonrotating
frame, but not a freely falling one. Here, we appreciate
that the force of gravity affects the optical properties of
the medium whereas the centrifugal force determines the
radius of the caustic circle and the diffraction pattern in
its neighborhood.
\par The time delay in radar sounding, the bending of light
in the vicinity of a massive body and the advance of the
perihelion will be treated from this viewpoint. The
introduction of the Newtonian potential will cause a
modification of the phase of the Bessel function and yield
both periodic and aperiodic orbits depending upon whether
the total energy is negative or positive, respectively. The
time delay in radar sounding is a double relativistic
effect that is caused by the slowing down of clocks in the
vicinity of a massive body, accounted for by the fact that
light travels at the phase velocity, and the shrinking of
measuring rods, accounted for by a line element that is
magnified by the index of refraction. Since an index of
refraction that takes into account  the Newtonian
gravitational potential only  gives  one half the
relativistic value it cannot be the cause, or at least the
entire cause, of the bending of light. Rather, if we take the
next higher interaction into account that causes
mass to accelerate we do  find the actual general
relativistic value without the Newtonian gravitational
potential. This implies that, like gravitational radiation,
the interaction between a light ray and a massive body is
predominantly quadrupole.\par In contrast, the advance  of
the perihelion requires both gravitational potential and the
quadrupole interaction. In general relativity, the quadrupole
appears as a relativistic correction to the square
of the transverse velocity in the conservation of energy. The
gravitational potential is responsible for the closed
elliptical  orbit, while the quadrupole causes the
perihelion to slowly rotate producing a rosette orbit. A
dipole moment would have been sufficient to cause the
advance of the perihelion
\cite{Born}, but, since there is conservation of momentum,
the center of mass of the system  cannot accelerate, and so
the mass dipole moment cannot accelerate.
\section{Metric and Phase}
   Fermat's principle states that the ray path connecting
two arbitrary points makes the optical path length
\begin{equation}
I=\int\,\eta\sqrt{2T}\,dt
\label{eq:Fermat}
\end{equation}stationary,   where $\eta$ is the
index of refraction and $T$ is the kinetic energy per unit
mass.    As a first application of Fermat's principle we
calculate the time delay in radar sounding. In general
relativity, this time delay is predicted by the Schwarzschild
metric evaluated on the null geodesic when all angular
dependencies are ignored. If a light signal is sent from the
earth, located along the
$x$-axis at $-x_E$,   to Venus,  which is located
behind the sun at $x_V$, the light ray will be bent
as it passes the gravitational field of the sun. Clocks will
be slowed down and the time it takes the ray to bounce off
Venus' surface and return to earth will be longer than if
the sun were not present. \par
The simplest mechanical analog of the index of refraction is
$\eta=\sqrt{1-2U/c^2}$, where $U$ is the potential energy
per unit mass, and $c$ is the speed of light in a vacuum.
Since the gravitational field of the sun   makes the medium
optically more dense, $U$ is identified as the gravitational
potential $-GM/r$, where
$G$ is the Newtonian gravitational constant, $M$ is the sun's
mass,  $r=\sqrt{R^2+x^2}$ is the distance from the center
of the sun to Venus,    and $R$ is the
radius of the sun.\par Now, according to Fermat's principle,
the propagation time $t$ along a ray connecting the two
endpoints
$-x_E$ and
$x_V$ is given by
\[t=\int_{-x_E}^{x_V}\,\frac{\eta(r)}{u(r)}\,dx=
\int_{-x_E}^{x_V}\,\frac{\sqrt{1+\mathcal{R}/r}}{u(r)}\,dx
  \] where $\mathcal{R}:=2GM/c^2$ is the Schwarzschild radius.
The slowing down of clocks in a gravitational field will
result in an apparent reduction in the speed of light.
Light will therefore travel at the phase velocity
$u(r)=c/\eta(r)$, rather than $c$, as it would in vacuum.
Consequently, the travel time will be
\[t+\Delta
t=t_N+\frac{\mathcal{R}}{c}\int_{-x_E}^{x_V}\,
\frac{dx}{r}=t_N+\frac{\mathcal{R}}{c}\int_{-x_E}^{x_V}\frac{dx}
{\sqrt{R^2+x^2}},\] where  $t_N=(x_V+x_E)/c$ is the
Newtonian travel time. The second term is half the time
delay for a signal to bounce off Venus and return to the
earth. Fermat's principle thus predicts a time dilatation of
\[\Delta t=\left(\frac{2\mathcal{R}}{c}\right)\ln\left(
\frac{x_V+\sqrt{R^2+x_V^2}}{-x_E+\sqrt{R^2+x_E^2}}\right)
\approx\left(\frac{2\mathcal{R}}{c}\right)\ln\left(
\frac{4x_Ex_V}{R^2}\right)=2.4\times 10^{-4} \;\mbox{s},\]
where the square roots have been expanded to lowest order
using the fact that $R\ll x_V, x_E$.\par This simple
application of Fermat's principle gives the same result as
general relativity,\footnote{In a simplified
demonstration of the time delay of radar sounding caused by a
massive body
\cite{Sexl},
$t=\int dx/c_{\mbox{eff}}$, where
$c_{\mbox{eff}}=
(1-\mathcal{R}/2r)^2c\approx(1-\mathcal{R}/r)c$,
is an \lq effective speed\rq\ that supposedly accounts for
both   time dilatation and the shrinking of measuring rods
in a gravitational field. The final expression for $t$ is
valid to first order in $\mathcal{R}/r$.  Rather, in our
presentation, the expression for $t$ is exact, and    it is
a result of two factors:  the slowing down of clocks, as
expressed by the fact that light rays propagate at the phase
velocity, $c/\eta(r)$, and the warping of measuring sticks,
as accounted for by the stretching of the line element,
$\eta(r)dx$. If the effective velocity
$c_{\mbox{eff}}$ were to be identified as the phase
velocity, then this would account for only half of the
effect.} and which is to  within
$3$
\% of experimental uncertainty \cite{Shapiro}.  In the
following, we will use gravitational units where $c=G=1$,
and the more general expression for the index of refraction
\begin{equation}
\eta=\sqrt{-A-2U},\label{eq:eta}
\end{equation} where the dimensionless  constant $-A$
represents the negative of twice the  total energy,
allowing for both positive and negative values.
   \par In a rotating coordinate system in the Euclidean plane
($\theta=\pi/2$)
\begin{equation}
2T=\dot{r}^2+r^2\dot{\phi}^2. \label{eq:T}
\end{equation}
   Fermat's
principle will determine the phase of the wave function
$\psi$  that is a solution of Helmholtz's equation,
\[
\left(\Delta+\kappa^2\right)\psi=0, \]
in the short-wavelength, or high-wave number
$\kappa$, limit, where $\Delta$ stands for the Laplacian. The
solution to Helmholtz's equation can be written as
\begin{equation}
\psi^{\pm}(r)=\frac{\mathcal{A}}{\sqrt[4]{\kappa^2
S^\prime}}e^{\pm i\left(\kappa
S-\vartheta\right)}\left(1+O(\kappa^{-1})\right),
\label{eq:psi}
\end{equation}
  where $\mathcal{A}$ is a constant term in the amplitude,
$\pm(\kappa S-\vartheta)$ are phases of the incoming and
outgoing waves, and $\vartheta$ is an integration constant
that will be determined by matching conditions between the
periodic and exponential domains [\textit{vid}.,
(\ref{eq:theta}) below].  The prime denotes differentiation
with respect to
$r$. The function
$S$ is referred to as the eikonal, and it is the object of
our study.
\par Introducing (\ref{eq:T}) into (\ref{eq:Fermat}) implies
that the ray path connecting two arbitrary points makes the
optical length
\begin{equation}
I= \int\,\eta\sqrt{\dot{r}^2+r^2\dot{\phi}^2}\,dt=
\int\,\eta\sqrt{1+r^2\phi^{\prime\,2}}\,dr
\label{eq:Fermat-bis}
\end{equation}
stationary. Observing that
$\phi$ is a cyclic coordinate, and calling $\Lambda$ the
integrand of (\ref{eq:Fermat-bis}), we immediately obtain a
first integral of the motion
\[
\frac{\partial\Lambda}{\partial\phi^{\prime}}
=\frac{\eta r^2\phi^{\prime}}
{\sqrt{1+r^2\phi^{\prime\,2}}}=r_a=\mbox{constant},
\]
regardless of whether the medium is homogeneous or not.
In an inhomogeneous medium $\eta$ will be a function
$r$. For the moment we shall assume that it is a
constant.  Solving for
$\phi^{\prime}$, we obtain the equation of the orbit as
\begin{equation}
d\phi=\frac{r_a\,dr}{r\sqrt{\left(\eta
r\right)^2-r_a^2}}.  \label{eq:orbit}
\end{equation}
Integration of
(\ref{eq:orbit})  gives the trajectory
\begin{equation}
\phi-\phi_0= \cos^{-1}\left(\frac{r_a}{\eta r}\right),
\label{eq:phi}
\end{equation}
where $\phi_0$ is a constant of integration.
  Squaring both sides of (\ref{eq:orbit}), and using
(\ref{eq:eta}) give the conservation of energy
\[
\dot{r}^2+r^2\dot{\phi}^2+2U(r)=-A, \]
provided the radius of the caustic \cite{Keller} is given by
\begin{equation}
r_a=r^2\dot{\phi}=\mbox{const}. \label{eq:ra}
\end{equation}
  This is   Kepler's law of equal
areas in equal times, or the conservation of angular
momentum, where
$r^2\dot{\phi}$ is the angular momentum (relative to unit
mass).   Introducing (\ref{eq:orbit}) into
(\ref{eq:Fermat-bis}) results in
\begin{equation}I=\eta\int\,\frac{rdr}{\sqrt{r^2-r_a^2/\eta^2}}
=\eta^2t=\sqrt{(\eta r)^2-r_a^2}.\label{eq:I}
\end{equation}
\par
The eikonal is the integral over $r$ of the Legendre
transform of the integrand of Fermat's principle
\begin{eqnarray}
S & = & \int\,\left(\Lambda-\phi^{\prime}
\frac{\partial\Lambda}{\partial\phi^{\prime}}\right)\,dr
=\int\,\frac{\sqrt{(\eta r)^2-r_a^2}}{r}\,dr\nonumber\\
& = & I-r_a(\phi-\phi_0)
= \sqrt{(\eta
r)^2-r_a^2}-r_a
\cos^{-1}\left(\frac{r_a}{\eta r}\right),\nonumber\\
& = & r_a\int\tan^2\phi\,d\phi=r_a(\tan\phi-\phi)\ge0.
\label{eq:S}
\end{eqnarray}
The second line of (\ref{eq:S}) expresses the eikonal as
   the difference between (\ref{eq:I}) and the arc
length along the caustic
\cite{Keller}. The third line of (\ref{eq:S}) shows that the
product of
$r_a$ and (\ref{eq:S}) is the area of a circular cap whose
base is a circular arc of radius $r_a$, and whose peak is a
distance $r$ from the center of the circle \cite{Sholander}.
In other words, the addition of a circular cap to a body
increases its area by $r_aS(r)$, while its perimeter is
increased by $2S(r)$,
which is strictly increasing and strictly convex on
$r_a\le r<\infty$.\par
\par The second line of (\ref{eq:S}) is Debye's
asymptotic expression for the phase of the Bessel function
in the periodic region
$r>r_a$ \cite{Debye}. In a homogeneous refractive body of a
constant index of refraction, a light source is situated a
distance
$r$ from the center of a circle of radius $r_a$, which
is the caustic. The eikonal (\ref{eq:S}) consists of two
rays: a half-line
$\sqrt{r^2-r_a^2}$ from the source to the point of tangency
to the circumference of a circle of radius $r=r_a$, and a ray
along the arc length, $r_a\cos^{-1}(r_a/r)$. This quantity
is subtracted from the straight line segment because the ray
is taken from the caustic to the source. For the reverse
path, the signs of the two terms are exchanged. Apart
from a term $-\fourth\pi$, expression (\ref{eq:S}), when
multiplied by $\kappa$, represents the phase going away from
the caustic, while the ray going toward the caustic has the
signs reversed in addition to a phase factor of
$+\fourth\pi$. Hence, the phase changes abruptly by an amount
$-\half\pi$ upon passing through the caustic. A phase jump
of this magnitude is observed for a focal line due to the
convergence of rays of a cylindrical wave
\cite{Sommerfeld}.\par
In the shadow region $r<r_a$,
where the rays do not penetrate, the eikonal (\ref{eq:S})
becomes completely imaginary
\begin{equation}
S_B^{\dagger}(r)=i\left\{r_a\cosh\left(\frac{r_a}{r}\right)-
\sqrt{r_a^2-r^2}\right\}.
\label{eq:S-hyper}
\end{equation}
Since the \lq shadow\rq\ intensities vanish rapidly as
$\kappa$, or the distance from the caustic increases, they
are usually ignored \cite{Poston}. However, the matching
conditions between the periodic and exponential regions are
of fundamental importance in quantum mechanics because they
furnish the quantum conditions \cite{Lanczos}. Hence we must
give credence to the type of motion that occurs in the shadow
zone. Interestingly enough, contradictions will arise with
the laws of physics that govern the illuminated, or
periodic, zone.\par For instance, by reinstating the velocity
of light
$c$, we have the inequality $r\dot{\phi}>c$ in the shadow
zone. This is in contradiction with special relativity.
However,  in the spectrum of a medium of anomalous
dispersion there can exist a region near the absorption line
where the group velocity can be greater than
$c$ \cite{Brillouin}. In this region, the group
velocity no longer represents the velocity of the
signal. Anomalous dispersion that results from strong
absorption destroys the characteristic wavelength of
propagation so that light can only be defined statistically.
For electrodynamic guided waves
\cite[p 149]{Brillouin}, and in quantum mechanics,   the
phase velocity   is greater than $c$. The concept of a
phase velocity in quantum mechanics was discarded when
Schr\"odinger [1982] proved that the group velocity of
the wave represented the particle velocity of the
electrons. But this meant that individual waves could not
be used for signal transmission. Even if they could be there
would still be a detection problem since no optical effect
could propagate with a velocity greater than
$c$.
\par
Boundary conditions in general relativity are usually stated
by requiring spacetime to be asymptotically flat, as
in the case of the Schwarzschild metric. But, in rotating
systems a cutoff must be introduced for otherwise distances
$r>c/\dot{\phi}$ would make the time component of the metric
tensor negative \cite{Landau}.  Such a restriction would
limit the phase of the Bessel function to the periodic
region, but it would not be in the form of an asymptotic
boundary condition. Larger distances where the angular
velocity becomes greater than the velocity of light would
bring us within the caustic region, and does not lead to the
conclusion that such a system cannot be made up of real
bodies \cite{Landau}.
\par The eikonal (\ref{eq:S-hyper}) can also be
derived from Fermat's principle, which now reads
\begin{equation}
I=\int\,\eta\sqrt{r^2\dot{\phi}^2-\dot{r}^2}\,dt=\int\,
\eta\sqrt{r^2\phi^{\prime\,2}-1}\,dr=\mbox{extremum}.
\label{eq:Fermat-hyper}
\end{equation}
Following the same procedure as before, we find the
trajectory
\begin{equation}
\phi-\phi_0=\cosh^{-1}\left(\frac{r_a}{\eta r}\right),
\label{eq:phi-hyper}
\end{equation}
for a constant index of refraction,  where we can always
arrange for $\phi_0=0$ by suitably fixing the initial point
for the measurement of the arc length. The extremum
(\ref{eq:Fermat-hyper}) is just the distance
$I=-\sqrt{r_a^2-(\eta r)^2}=-r_a\tanh\phi$. \par Using
the canonical parameterization  for which
$r_a=1$, the arc length $s=\sinh\phi$ enables the profile
curve to be written as
\[\beta(s)=(g(s),h(s))=\left(\sinh^{-1}s-
\frac{s}{\sqrt{1+s^2}},\frac{1}{\sqrt{1+s^2}}\right),\]
where $g(s)$ measures the distance along the axis of
revolution, and $h(s)$ measures the distance from the axis
of revolution. The parameterization of the surface of
revolution is
\[x(s,\theta)=\left(\sinh^{-1}s-\frac{s}{\sqrt{1+s^2}},
\frac{\cos\theta}{\sqrt{1+s^2}},\frac{\sin\theta}
{\sqrt{1+s^3}}\right),\]
where $\theta$ is the angle through which the profile
curve has been rotated. The element of arc length on the
surface of revolution is
\[d\sigma^2=dr^2+r^2\mbox{sech}^2\phi\left(\tanh^2\phi\,d\phi^2
+d\theta^2\right).\]\par
The eikonal (\ref{eq:S-hyper}) is just the distance along
the $\phi$ axis,
\[
S^{\dagger}_B(s)=\sinh^{-1}s-\frac{s}{\sqrt{1+s^2}}
\] which in terms of $\phi$,
\begin{equation}
S_B^{\dagger}(\phi)=
r_a\int\,\tanh^2\phi\,d\phi=r_a(\phi-\tanh\phi).
\label{eq:S-hyper-bis}\end{equation}
is a tractrix, having a constant negative
curvature, $K=-1/r_a^2$.
The tangent to the tractrix which intersects the
$x$-axis always has the constant value $r_a$. The distance
from the origin to the point of tangency is
$r_a\phi$. The point  on the tractrix which has a tangent
intercepting the $x$-axis is located a distance
$r_a(\phi-\tanh\phi)$ along the $x$-axis. This is precisely
the eikonal (\ref{eq:S-hyper-bis}). Hence, in the periodic
domain the eikonal is half the increase in the perimeter due
to the addition of a spherical cap to a body, while in the
exponential domain, the eikonal is a tractrix, which is the
involute of a catenary unwinding from its lowest point.
\par
The solutions to Helmholtz's equation are now exponentials
\[
\psi^{\pm}(r)=\frac{\mathcal{B}^{\pm}}{\sqrt[4]{\kappa
S^{\dagger\,\prime}}} e^{\pm\kappa
S^{\dagger}}\left(1+O(\kappa^{-1})\right).
\]
The general solution has two free constants,
$\mathcal{B}^{+}$ and
$\mathcal{B}^{-}$, associated with the $\pm$ signs in the
exponent. The exponential growing solution is unphysical, and
consequently, it must vanish. The matching condition between
the constants
$\mathcal{A}$ and
$\vartheta$ in (\ref{eq:psi}) and $\mathcal{B}^{+}$ is
\cite{Lanczos}
\[\mathcal{B}^{+}=\half
\mathcal{A}\cos(\vartheta-\fourth\pi),\] and if it is to
vanish,
\begin{equation}
\vartheta=-\fourth\pi\pm k\pi, \label{eq:theta}
\end{equation}
where $k$ is an arbitrary integer. This gives rise to the
phase jump $e^{i\fourth\pi}$ as the system passes through
the caustic.
\section{The Mechanics of Diffraction Phenomena}
Formulas (\ref{eq:ra}) and (\ref{eq:I}) are familiar from
Kepler's theory, but depend on the form of the index of
refraction. In a repulsive field $U$ is positive, or in no
field at all, the realness of the index of refraction
(\ref{eq:eta}) requires $A<0$ so that the trajectory is
open, extending between $r=\infty$ and a minimum value of
$r$, where the velocity vanishes.\par In an attractive field
of force,
$A$ can be of either sign. The unperturbed motion will be
the result of some playoff between the gravitational
potential, $U=\mathcal{R}/2r$ and the centrifugal potential,
$r_a^2/2r^2$. Introducing the former into the expression for
the index of refraction (\ref{eq:eta}), the eikonal
(\ref{eq:S}) becomes
\begin{eqnarray}
S(r,r_a) & = & \int\,\frac{\sqrt{-Ar^2+\mathcal{R}r-r_a^2}}
{r}\,dr\nonumber\\
& = & \sqrt{-Ar^2+\mathcal{R}r-r_a^2}-r_a\cos^{-1}
\left(\frac{q/r-1}{\epsilon}\right)\nonumber\\
& & +\half\mathcal{R}\left\{
\begin{array}{c}
\frac{1}{\sqrt{A}}\cos^{-1}\left(\frac{1-r/a}{\epsilon}
\right)\label{eq:S-bis}\\
\frac{1}{\sqrt{-A}}\cosh^{-1}\left(\frac{1+r/a}{\epsilon}
\right)
\end{array} \right.
\end{eqnarray}
  where the eccentricity
\begin{equation}
\epsilon=\sqrt{1-\frac{4Ar_a^2}{\mathcal{R}^2}},
\label{eq:e}
\end{equation}
the semi-latus rectum
\begin{equation}
q=\frac{2r_a^2}{\mathcal{R}}=a\left(1-\epsilon^2\right),
\label{eq:q}
\end{equation}
and $a=\mathcal{R}/2|A|$, which is the semi-major axis if
$\epsilon<1$. The equation for the orbit is
\begin{equation}
r=\frac{q}{1+\epsilon\cos\phi} \label{eq:ellipse}
\end{equation}
which is an ellipse for $\epsilon<1$ ($A>0$), or a
hyperbola for $\epsilon>1$ ($A<0$). In the periodic case,
$\phi$ is known as the \lq true anomaly\rq, and $u$ is the
\lq eccentric anomaly\rq.  The latter satisfies the equation
\[
r=a\left(1-\epsilon\cos u\right) \]
in the periodic case, $A>0$, while it satisfies
\[r=a\left(\epsilon\cosh u-1\right),\]
in the hyperbolic case $A<0$.\par
The velocity
\begin{equation}
\dot{r}=\pm\frac{\sqrt{-Ar^2+ \mathcal{R}r-r_a^2}}{r},
\label{eq:r-dot}
\end{equation}
will vanish at
\begin{equation}
r_{\pm}=a(1\pm\epsilon). \label{eq:aphelion}
\end{equation}
$r$ undergoes a libration, where the turning points are the
aphelion ($r_+$) and  the perihelion ($r_-$). In the
hyperbolic case, there will be only one point where the
velocity (\ref{eq:r-dot}) vanishes, and that is at the
closest distance of approach
\begin{equation}
r_{\min}=a(\epsilon-1). \label{eq:r-min}
\end{equation}
\par
A Hankel function of type $j$, with argument $\kappa\epsilon
r$, and order $\kappa(q-r)$ has the integral representation
\begin{equation}
H^{j}_{\kappa(q-r)}(\kappa\epsilon r)=\frac{1}{\pi}
\int_{C_{j}}e^{i\kappa\left(\epsilon r\sin\phi-(q-r)\phi
\right)}\,d\phi,\;\;\;\;\;\;\;\; j=1,2. \label{eq:Hankel}
\end{equation}
The contours $C_j$ are the paths from $-\pi+i\infty$
to $-i\infty$ and $-i\infty$ to $\pi+i\infty$.   We seek an
asymptotic expansion of (\ref{eq:Hankel}) as
$\kappa\rightarrow\infty$. On the basis of the definitions of
the contours $C_j$, the only critical points of
\begin{equation}
W(\phi)= \epsilon r\sin\phi-(q-r)\phi
\label{eq:W}
\end{equation}
are simple saddle points. The necessary condition for
simple saddle points,
$W^{\prime}(\phi_{\pm})=0$, gives   equation
(\ref{eq:ellipse}), where $\phi_{+}=-\phi_{-}$ with
$0<\phi_{+}<\pi$ and $-\pi<\phi_{-}<0$. These saddle points
will be simple because
$W^{\prime\prime}(\phi_{\pm})=-\epsilon
r\sin\phi_{\pm}\neq0$.   Evaluating
(\ref{eq:W}) at the saddle points leads to
\begin{equation}
W(\phi_{\pm})=\pm\left\{\frac{q}{r_a}
  \sqrt{-Ar^2+\mathcal{R}r-r_a^2}-\left(q-
r\right)\cos^{-1}\left(
\frac{q-r}{\epsilon r}\right)\right\}. \label{eq:W-bis}
\end{equation}
As $\epsilon\rightarrow\infty$, implying that $A<0$ and the
path is an hyperbola, (\ref{eq:W-bis}) becomes proportional
to the free eikonal (\ref{eq:S}). The derivative of
(\ref{eq:W-bis}) with respect to $r$ is proportional to the
velocity (\ref{eq:r-dot}) in this limit. We will have need
of this limit in the next section when dealing with the
bending of light by a massive body.
\par
If $q-r>\epsilon r$, the order of the Hankel function
(\ref{eq:Hankel}) is greater than its argument. The
  saddle points $\phi_{\pm}$ now lie on the
imaginary axis
\[\phi_{\pm}=\pm i\cosh^{-1}\left(\frac{q-r}{\epsilon r}
\right),\]
and the phase becomes completely imaginary
\[W^{\dagger}(\phi_{\pm})=\pm
i\left\{\sqrt{(q-r)^2-\epsilon^2r^2}- (q-r)
\cosh^{-1}\left(\frac{q-r}{\epsilon
r}
\right)\right\}
  .\]
This corresponds to the shadow region where the rays do not
penetrate. Since the contributions from $\phi=\phi_{-}$ are
exponentially smaller than those from $\phi=\phi_{+}$, the
former can be neglected; hence, to leading order
\begin{eqnarray*}\lefteqn{H^{(j)}_{\kappa(q-r)}(\kappa\epsilon
r)  \sim  \sqrt{\frac{2}{\pi\kappa}}\times}\\ & &
\frac{\exp\left\{\kappa\left[(q-r)\cosh^{-1}((q/r-1)
/\epsilon)-(q/r_a)\sqrt{Ar^2-\mathcal{R}r+r_a^2}\right]+
(-1)^{j}i\half\pi\right\}}
{\sqrt[4]{Ar^2-\mathcal{R}r+r_a^2}},
\end{eqnarray*}
for $j=1,2$. The change of signs under the square root
sign imply that the gravitational potential is now
repulsive, while the centrifugal potential is attractive.  In
the periodic case, the shadow region lies beyond the limits
of libration (\ref{eq:aphelion}), whereas in the aperiodic
case, it is a circle whose radius is smaller than
(\ref{eq:r-min}). Mechanically, these regions are classically
inaccessible while optically, they can be penetrated, like
quantum mechanical tunneling, and the breakdown of
structural stability in catastrophe theory. Near a caustic
of the free phase Bessel function, it would be approximated
by a cubic resulting in an Airy function. The catastrophe   is
known as a fold catastrophe \cite{Poston}.
\par Finally, in
the case
$A=0$,    corresponding to a parabolic orbit
($\epsilon=1$), the eikonal is
\begin{eqnarray}
S(r,r_a) & = & \int\,\frac{\sqrt{\mathcal{R}r-r_a^2}}{r}\,dr=
2\sqrt{\mathcal{R}r-r_a^2} -
2r_a\cos^{-1}\left(\frac{r_a}{\sqrt{\mathcal{R}r}}\right)\nonumber\\
&  = &
2r_a\left(\tan\phi-\phi\right)\ge0,
\label{eq:S-tris}
\end{eqnarray}
where $\phi=\cos^{-1}\left(r_a/\sqrt{\mathcal{R}r}\right)$.
The eikonal (\ref{eq:S-tris}) is similar to the unbounded
motion, periodic motion of (\ref{eq:S}), with the
exception that the caustic radius $r_a$ is magnified by a
factor $r_a/\mathcal{R}$.\par

\section{Deflection of Light}

If $U$ is the gravitational potential, the interaction is
attractive, and the orbit curves towards the center of
force. Gravity makes the medium optically more dense in the
vicinity of the sun    than far away
from it. As a result, light waves will be bent around the
sun rather than being straight. The effect was originally
predicted by S\"oldner in 1801, and rederived by Einstein in
1911 on the basis of the Doppler effect and Newtonian theory.
Specifically, Einstein considered the slowing down of
light in a gravitational field \cite{Einstein}.  General
relativity predicts a value twice as great, as Einstein
showed in 1915
\cite{Einstein}. It is therefore concluded that Newtonian
theory is only an approximation, valid at speeds small
compared to that of light, and the full theory must be used
to calculate the actual deflection of a light beam traveling
through the sun's gravitational field. The value
obtained from Newtonian theory is only half that predicted
from general relativity. If instead the interaction
were quadrupole in nature, the phase of the Bessel
function would belong to the periodic domain, and thus
correspond to a diffraction phenomenon.\par
The directions of the asymptotes of the hyperbola
(\ref{eq:ellipse}) are determined from the condition
$r=\infty$ which gives
\begin{equation}
\Delta\phi=\cos^{-1}\left(-\frac{1}{\epsilon}\right)=
\pi-\cos^{-1}\left(\frac{1}{\epsilon}\right)=\half\pi+
\sin^{-1}\left(\frac{1}{\epsilon}\right). \label{eq:bend}
\end{equation}
Since only the rest energy is involved, $A=-1$, and
$\epsilon=2r_a/\mathcal{R}$ for $r_a\gg\mathcal{R}$, the
total deflection will be twice (\ref{eq:bend}) or
\[2\Delta\phi=\pi+2\sin^{-1}
\left(\frac{\mathcal{R}}{2r_a}\right).\]
A straight ray corresponds to  $2\Delta\phi=\pi$.
  This is   comparable to Coulomb
scattering, except that the trajectory is concave toward the
origin where the massive body is located. For small angles of
deflection the angle between the two asymptotes differs from
$\pi$ by
$\theta=\mathcal{R}/r_a$,
which is the Newtonian result, but
differs from the actual gravitational value by a factor of
two. This is  used to highlight the shortcomings of
Newtonian theory and the need of general
relativity \cite{Moller}. \par
M\o ller [1952] splits the deflection of light into two
effects: a velocity of light that varies as
$1/\sqrt{1-\mathcal{R}/r}$, and the non-Euclidean character
of the spatial geometry. The first effect gives  the
radial velocity
\begin{equation}
\dot{r}=\sqrt{\frac{1}{1-\mathcal{R}/r}-\frac{r_a^2}{r^2}}
=(\ref{eq:r-dot}),\label{eq:r1}
\end{equation}
to first order in $\mathcal{R}/r$, while the second effect
has a radial velocity
\begin{equation}
\dot{r}=\sqrt{1-\frac{\mathcal{R}}{r}-\frac{r_a^2}{r}
+\frac{\mathcal{R}r_a^2}{r^3}}.\label{eq:r2}
\end{equation}
Both give a contribution that is half the general
relativistic value.   Equation (\ref{eq:r2}), which arises
from the Schwarzschild exterior metric tensor, has the wrong
sign of gravitational field. It appears repulsive like that
of the centrifugal potential. Rather, we shall now
show that the deflection of light by a massive body is
accountable by the last term in (\ref{eq:r2}), which is a
quadrupole interaction,
\textit{without} the incorrect gravitational potential.
\par
When the light ray enters the scattering potential, it
enters a region of changing index of refraction, where it is
both refracted and diffracted. This is determined by the
phase of the wave function.  Since the gravitational
potential is not responsible for the bending of light, we
have to look to a higher order interaction. Expanding the
gravitational energy of the central field of force in
inverse powers of
$r$ we have
\begin{equation}
2U=-\frac{\mathcal{R}}{r}\left(1+c_1\frac{r_a}{r}+
c_2\left(\frac{r_a}{r}\right)^2+\cdots\right),
\label{eq:U}
\end{equation}
where the $c_i$ are coupling constants, and $r_a$ is
the characteristic length. This is analogous to the Coulomb
case, where $r_a$ would represent the radius of the first
Bohr orbit \cite{Sommerfeld-bis}.  The next lowest
interaction, $c_1$, would be a mass dipole which, however,
cannot oscillate so that its center of mass cannot
accelerate. Hence, we set
$c_1=0$, and consider the quadrupole term,
$\mathcal{Q}/r^3$, where $\mathcal{Q}=Mr_a^2$ is the
quadrupole moment of a system consisting of the sun and a
photon in the direction
  perpendicular to the asymptote of the photon
trajectory and the parallel line passing through the center
of the sun. We shall consider the coupling constant
$c_2$ to be of order unity.\par It proves convenient
\cite[p 354]{Moller} to introduce the new variable
$\sigma=r_a\rho\sqrt{1-\mathcal{R}\rho}$ into the
equation of the trajectory,
\begin{equation}
\frac{d\rho}{d\phi}=\pm\sqrt{\frac{1}{r_a^2}-\rho^2+
\mathcal{R}\rho^3}.\label{eq:quad}
\end{equation}
Neglecting the small term $\mathcal{R}\rho^3$, the equation
of the orbit is (\ref{eq:phi}), which is equivalent to
\[\phi+\frac{\partial S}{\partial r_a}=\phi_0=\mbox{const.},
\]
where $S$ is the eikonal of the Bessel function,
(\ref{eq:S}). For a constant index of refraction, the
trajectory $r=r_a/\eta\sin\phi$, obtained by setting
$\phi_0=\half\pi$, is a straight line which passes through
the origin at a distance $r_a/\eta$ when $\phi=\half\pi$ and
goes to infinity again for $\phi\rightarrow\pi$.
\par
The exact equation (\ref{eq:quad}) may be written as \cite[p
354]{Moller}
\begin{equation}
\frac{d\rho}{d\phi}=\frac{1}{r_a}
\sqrt{1-\sigma^2}, \label{eq:Moller}
\end{equation}
where $\sigma=r_a\rho\sqrt{1-\mathcal{R}\rho}$. Since
$\mathcal{R}\rho$ is a small quantity,
  we obtain  the approximations
\[\sigma=r_a\rho\left(1-\half\mathcal{R}\rho\right)
\hspace{20pt}r_a\rho=\sigma\left(1+\half\mathcal{R}\rho
\right)=\sigma\left(1+\frac{\mathcal{R}}{2r_a}\sigma\right),\]
to first order, and consequently,
$r_ad\rho=(1+\mathcal{R}\sigma/r_a)d\sigma$. Introducing
these approximations into (\ref{eq:Moller}), and integrating
from $0$ to  the distance of closest approach, $\sigma_0$, we
get
\begin{equation}
\Delta\phi=\pi+\theta=2\int_0^{\sigma_0}
\frac{d\sigma(1+\mathcal{R}\sigma
/r_a)}{\sqrt{1-\sigma^2}}=\pi+2\frac{\mathcal{R}}{r_a}.
\label{eq:phi-bis}
\end{equation}
In the derivation of (\ref{eq:phi-bis}) we have used the
fact that the distance of closest approach is determined by
the vanishing of the integrand in the denominator, and,
hence,
$\sigma_0=1$. Consequently, the deflection
$\theta=2\mathcal{R}/r_a$ is twice that obtained by
treating the interaction through a Newtonian potential. The
quadrupole interaction is introduced by the  index of
refraction, while the geometry is determined by the flat
metric, proportional to (\ref{eq:T}). Can we associate a
wave function to such a process?
\par
For the quadrupole interaction we have a
wave function whose eikonal and amplitude are approximately
given by
\begin{equation}
S_Q(r)=
\sqrt{r^2-r_a^2\left(1-
\frac{\mathcal{R}}{r}\right)}\\
-r_a\cos^{-1}\left(\frac{r_a}{r}\sqrt{1-\frac{\mathcal{R}}{r}}
\right),
\end{equation}
and
\begin{equation}
\mathcal{A}\Bigg/\sqrt[4]{1-\frac{r_a^2}{r^2}\left(
1-\frac{\mathcal{R}}{r}\right)},\label{eq:A}
\end{equation}
respectively, since $\mathcal{R}/r$ is a very small
quantity. The square of the amplitude, (\ref{eq:A}), is
an intensity which measures the \lq geometric divergence\rq\
of the wave field
\cite{Babic}. The larger it is, the more the rays in the
neighborhood of some fixed ray will diverge from each other.
In the absence of the quadrupole interaction the geometric
divergence becomes infinite on the caustic $r=r_a$, which is
the case of the ordinary Bessel function. In the presence of
the quadrupole term, the geometric divergence of the
ray field becomes infinite at a distance smaller than
$r_a$. The equation of the trajectory,
\[r=\frac{r_a}{\sin\phi}\sqrt{1-\frac{\mathcal{R}}{r}},\]
shows that the trajectory becomes infinite as
$\phi\rightarrow0$ and $\pi$,  while the distance of closest
approach
$r_0$ is less than the caustic radius $r_a$. Consequently, as
the result of the quadrupole interaction the distance from
the scattering center to the caustic has been reduced. In the
limit as
$r\rightarrow r_0$, the form of the wave function
(\ref{eq:psi}) is no longer valid, and a uniform asymptotic
expansion of the Bessel function can be given in terms of
Airy functions \cite{Ludwig}.\par
The differential cross section of the quadrupole interaction
is
\[\Delta(\theta)=2\pi r_a\Big|\frac{dr_a}{d\theta}\Big|=
8\pi\frac{\mathcal{R}^2}{\theta^3},\]
which displays the characteristic scattering property of the
cross section to approach infinity as the angle of
deflection approaches zero.  However, since the rest energy
is involved, there will be no rapid decrease of $\theta$
with the kinetic energy, as is usually the case in particle
scattering.
\section{The Perihelion Shift}
Thus far we have treated open trajectories $A<0$. In the case
of Mercury, we are dealing with a closed orbit, and hence
negative total energy $A>0$, which is not exactly elliptical
due to the advance of the perihelion. This precession
amounts to
$43^{\prime\prime}$   in a century, and was known to
astronomers as early as 1860. However, its explanation had
to await the general theory of relativity in 1915.\par
Taking into account both the gravitational potential and the
quadrupole interaction, the index of refraction is
[\textit{vid}. (\ref{eq:U}) with
$c_1=0$ and
$c_2=1$]
\[
\eta= \sqrt{-A+\frac{\mathcal{R}}
{r}\left(1+\frac{r_a^2}{r^2}
\right)}. \]
  In the
unperturbed state, where the quadrupole interaction is
absent, the gravitational potential must be large enough so
that the eccentricity is real [\textit{vid}. (\ref{eq:e})
below].
  The eikonal is
\begin{equation}
S_P(r,r_a)   =
\int\,\frac{\sqrt{\eta^2r^2-r_a^2}}{r}\,dr
   =  \int\,\frac{\sqrt{-Ar^2+\mathcal{R}r-r_a^2+\mathcal{R}
r_a^2/r}}{r}\,dr. \label{eq:SP}
\end{equation}
A closed trajectory will result from a dynamic balance
between   gravitational  and   centrifugal forces.\par
In the presence of the quadrupole interaction, the angular
momentum will exist only for radii
$r>\threehalves\mathcal{R}$. Introducing this fact into
(\ref{eq:SP}) through the transformation,
$r+\threehalves\mathcal{R}=r^{\prime}$, and retaining
terms that are at most quadratic in the Schwarzschild
radius, give
\[
S_P(r,r_a)\simeq\int\,\sqrt{-A+\frac{\mathcal{R}}{r}
\left(1+\threehalves\frac{\mathcal{R}}{r}\right)-
\frac{r_a^2}{r^2}\left(1+2\frac{\mathcal{R}}{r}\right)-
\ninehalves\frac{\mathcal{R}^2r_a^2}{r^4}}\,dr,
\]
where for brevity we have dropped the prime on $r$.
Expanding the integrand in powers of the small
correction terms results in
\begin{eqnarray}
S_P(r,r_a) & \simeq &
  S^{(0)}_P(r,r_a)+\threefourths
\mathcal{R}^2\int\,
\frac{dr}{r\sqrt{-Ar^2+\mathcal{R}r-r_a^2}}\nonumber\\
& & -\mathcal{R}r_a^2\int\,\frac{dr}{r^2
\sqrt{-Ar^2+\mathcal{R}r-r_a^2}}+\cdots \label{eq:SP-exp}
\end{eqnarray}
where the unperturbed eikonal 
$S^{(0)}_P(r,r_a)$ is given by (\ref{eq:S-bis}).
The unperturbed orbit,
$-S^{(0)}_P/\partial r_a=\phi$ is given by the ellipse
(\ref{eq:ellipse}) since $\epsilon<1$.\par
The change in
$S^{(0)}_P$ over one complete orbit is
\[\frac{\partial\Delta S^{(0)}_P}{\partial r_a}=2\pi.\]
  As $r$ goes through one libration, the
true anomaly $\phi$ increases by $2\pi$, and there is no
perihelion advance.  However, taking into
consideration the first order correction term in
(\ref{eq:SP-exp}), which can be written as
\[\Delta
S^{(1)}_P(r,r_a)=-\threefourths\frac{\mathcal{R}^2}{r_a}
\frac{\partial\Delta S^{(0)}_P}{\partial r_a}=
\threehalves\pi\frac{\mathcal{R}^2}{r_a},\]
and differentiating it with respect to $r_a$, give the
first order correction term
\begin{equation}\Delta\phi_1=-\frac{\partial\Delta S^{(1)}}
{\partial
r_a}=\threehalves\pi\frac{\mathcal{R}^2}{r_a^2}.
\label{eq:D-phi}
\end{equation}
Introducing the semi-latus rectum (\ref{eq:q}) results in
\begin{equation}
\Delta\phi_1=3\pi\frac{\mathcal{R}}{a(1-\epsilon^2)},
\label{eq:peri}
\end{equation}
which is the general relativistic result for the perihelion
shift. For Mercury, the rotation of the perihelion per
revolution amounts to $0.104^{\prime\prime}$. The
dimensionless energy constant $A=2.59\times 10^{-8}$, and
the mean motion
$\omega=r_ac/ab=2cA^{\threehalves}/\mathcal{R}=8.34\times
10^{-7}$ s$^{-1}$, where we have reinstated the speed of
light $c$, and $b=r_a/\sqrt{A}$ is the semi-minor axis. The
period of the motion of Mercury is
$T=2\pi/\omega=87.25$ days, which is close to the actual
value of $88$ days. The frequency of rotation of the perihelion will be
$\Delta\omega=\omega\Delta\phi_1=4.25\times 10^{-13}$
s$^{-1}$.
\par
There are five experimental confirmations of general
relativity: time delay in radar sounding,   deflection of
light,   perihelion advance, spectral shift, and the
geodesic effect. We have shown that three of these effects
can be treated as diffraction phenomena on the basis of
Fermat's principle and the modification of the phase of a
Bessel function in the short-wavelength limit. The
spectral shift can also be derived from energy conservation
by assuming that a photon's energy has both inertial and
gravitational mass. This depends in an essential way on the
equivalence principle, and   not upon general
relativity.
  

\begin{thebibliography}{Dillo 83}
\bibitem[Babi\v{c}, Buldyrev 1991]{Babic}Babi\v{c} V M and
Buldyrev V S (1991)
\textit{Short-Wavelength Diffraction Theory\/}. Berlin,
Springer, p 24
\bibitem[Born 1925]{Born}Born M (1925) \textit{Vorlesungen
\"uber Atommechanik\/}. Berlin, Springer, p 159
\bibitem[Brillouin 1960]{Brillouin}Brillouin L (1960)
\textit{Wave Propagation and Group Velocity\/}. New York,
Academic Press, pp 21--22
\bibitem[Debye 1988]{Debye}Debye P (1988) in \textit{The
Collected Papers of Peter J W Debye\/}. Woodbridge CT, Ox Bow
Press, pp 583--607
\bibitem[Einstein 1952]{Einstein}Einstein A (1952) in
\textit{The Principle of Relativity\/}. New York, Dover, pp
99, 111
\bibitem[Fock 1966]{Fock}Fock V (1966) \textit{The Theory of
Space, Time and Gravitation\/}. 2nd edn, Oxford, Pergamon, p 222
\bibitem[Keller, Rubinow 1960]{Keller}Keller J B and Rubinow
S I (1960)
\textit{Ann. Phys\/} (NY). \textbf{9}: 24--75
\bibitem[Lanczos 1961]{Lanczos}Lanczos C (1961)
\textit{Linear Differential Operators\/}. London, D. Van
Nostrand, p 393
\bibitem[Landau and Lifshitz 1975]{Landau}Landau L D and
Lifshitz E M  (1975) \textit{The Classical Theory of Fields\/}.
4th edn, Oxford, Pergamon, p 254
\bibitem[Ludwig 1970]{Ludwig}Ludwig D (1970) \textit{SIAM
Rev\/}. \textbf{12}: 325--221
\bibitem[M{\o}ller 1952]{Moller}M{\o}ller C (1952)
\textit{The Theory of Relativity\/}. 1st edn, London, Oxford U.
P., p 355
\bibitem[Poston and Stewart 1978]{Poston}Poston T and
Stewart I (1978) \textit{Catastrophe Theory and its
Application\/}. London, Pitman, p 264
\bibitem[Schr\"odinger 1982]{Schrodinger}Schr\"odinger E
(1982) \textit{Collected Papers on Wave Mechanics\/}. New York,
Chelsea, p 17
\bibitem[Sexl and Sexl 1979]{Sexl}Sexl R and Sexl A (1979)
\textit{White Dwarfs--Black Holes\/}. New York, Academic, pp
41--44
\bibitem[Shapiro 1968]{Shapiro}Shapiro I I
(1968) \textit{Phys. Rev. Lett\/}. \textbf{20}: 1265--1269;
Anderson J D, Esposito P B, Martin W, and Thornton C L (1975)
\textit{Astrophys. J\/}. \textbf{200}: 221--233
\bibitem[Sholander 1952]{Sholander}Sholander M (1952)
\textit{Trans. Amer. Math. Soc\/}. \textbf{23}: 139--173
\bibitem[Sommerfeld 1950]{Sommerfeld-bis}Sommerfeld A (1950)
\textit{Atombau und Spektrallinien\/}. 7th edn, Braunschweig, F.
Vierweg \& Sohn,  p 712
\bibitem[Sommerfeld 1964]{Sommerfeld}Sommerfeld A (1964)
\textit{Optics\/}. New York, Academic Press, p 318




\end{thebibliography}
\end{document}